\documentclass[preprint,showpacs,preprintnumbers,amsmath,amssymb]{revtex4}
\usepackage{graphicx}
\begin{document}
%\preprint{APS/123-QED}

%
% title page
%
\title{Synchronized pulse control of decoherence}  
\author{Chikako Uchiyama}
\affiliation{%
Faculty of Engineering, University of Yamanashi,\\
4-3-11, Takeda, Kofu, Yamanashi 400-8511, JAPAN
}%
\author{Masaki Aihara}
\affiliation{%
Graduate School of Materials Science,
Nara Institute of Science and Technology,\\
8916-5, Takayama-cho, Ikoma, Nara 630-0101 JAPAN
}%
%\date{\today} 
\date{September 17,2003}  
\def\ch{{\cal H}}
\def\hbar{\mathchar'26\mkern -9muh}
\def\muv{{\vec \mu}}
\def\ev{{\vec E}}
\def\ih{\frac{i}{\hbar}}
\def\taus{{\tilde \tau}_{s}}
\def\tauc{{\tilde \tau}_{c}}
\def\ee{|e\rangle \langle e|}
\def\eg{|e\rangle \langle g|}
\def\ge{|g\rangle \langle e|}
\def\gg{|g\rangle \langle g|}

%%%%%%%%%%% abstract %%%%%%%%%%%
\begin{abstract}
We present a new strategy for multipulse control over decoherence. When a two-level
system interacts with a reservoir characterized by a specific frequency, we find that the decoherence is effectively suppressed by synchronizing the pulse-train application with the
dynamical motion of the reservoir.  We discuss the applicability of this strategy by studying the dependence of the decoherence suppression on the shape of the coupling spectral density. 
We find that the effectiveness of this strategy arises from the non-Markovian nature of dynamical motion of the reservoir.  
\end{abstract}
%%%%%%%%%%%%%%%%%%%%%%%%%%%%%

\maketitle
%%%%%%%%%%% sec.1 %%%%%%%%%%%
\section{Introduction}
\label{sec:1}
%%%%%%%%%%%%%%%%%%%%%%%%%%%%%
Degradation of quantum superposed state by decoherence is an obstacle to quantum information processing. In order to proactively prevent errors, a multipulse control method has been proposed \cite{lloyd1,ban,luming}.  It is essential that the application of \(\pi\) pulses causes time reversal in order to suppress the decoherence. The method has attracted considerable attention; it has been applied to suppress unwanted spontaneous emission\cite{agarwal1,agarwal2}, the magnetic-state decoherence by collisions in a vapor\cite{search1,search2} and the damping of vibrational mode of a chain of trapped ions\cite{vitalia,vitalib}.  While the multipulse control method requires no ancillary bits and no accurate detection, its effectiveness has been shown when sufficiently short and strong pulses are periodically applied in a shorter interval than the characteristic time of the system-reservoir interaction\cite{lloyda,lloydb}.  The degree of suppression becomes larger as the pulse interval becomes shorter.  Since these conditions are not easy to execute, a new approach to use a continuous control field instead of pulses has been proposed \cite{viola2}.  It is also shown that the control pulses do not always have to be ultra short for systems coupled to the reservoir with \(1/f\) spectral density\cite{lidars}.  In order to overcome the strict condition on the pulse application, it is desirable to seek a possibility to use the pulse trains with relatively long pulse interval.  

In previous paper\cite{uchiyama}, we have suggested a possibility to relieve the condition on pulse interval by formulating a theory of pulse control on the pure dephasing phenomena that is caused by the interaction with a boson reservoir.  Since the ordinary spin-boson model where a spin linearly interacts with the boson reservoir cannot describe the irreversibility in the long time region except for the ohmic dissipation case, we have extended the model to include a nonlinear interaction.  We have found that the multipulse control is effective for this model when the pulse interval is shorter than the reservoir correlation time.  We also found that the effective pure dephasing time shows a non-monotonic dependence on the pulse interval, that is, it has a peak when an application of \(\pi\) pulse-train is synchronized with the oscillation of the reservoir.  This means that the pure dephasing phenomenon is also effectively suppressed by paying attention to the dynamics of the reservoir.  However, in many cases, the system-reservoir interaction is described with a strong linear interaction and a weak nonlinear interaction. In the time region where we want to discuss the effectiveness of suppression by synchronizing the pulse application and the dynamics of the reservoir, the linear interaction plays an essential role in the decay.      
  
In this paper, assuming that a two-level system linearly interacts with a boson reservoir that has a characteristic frequency, we discuss the effectiveness of the synchronization of a \(\pi\) pulse train with the oscillation of the reservoir.  For convenience, we name this strategy as synchronized pulse control(SPC) in the following discussion.  As recognized in \cite{lidars}, SPC also depends on the type of coupling(bath) spectral density.  In order to make clear the applicability of SPC, we study the effectiveness of SPC on non-Lorentzian and Lorentzian coupling spectral density.     

The outline of this paper is as follows: In Sec.\ref{sec:2}, we introduce the model of the decoherence and derive the basic formula for multipulse control on the linear spin-boson model.  Next, we discuss the synchronized pulse control in Sec.\ref{sec:3}: The application of the basic formula to non-Lorentzian (Lorentzian) coupling spectral density is written in Sec.\ref{sec:3}A(B), respectively.  After discussing the effectiveness of the SPC in Sec.\ref{sec:4}, we give concluding remarks in Sec.\ref{sec:5}.

%%%%%%%%%%% sec.2 %%%%%%%%%%%
\section{Formulation}
\label{sec:2} 
%%%%%%%%%%%%%%%%%%%%%%%%%%%%%
We consider a two-level system composed of an excited state \(|e\rangle\) and a ground state \(|g\rangle\) with energy \(E_e\).  Let us consider the decoherence of this two-level system, which is caused by a linear interaction between the excited state and a boson reservoir. The system Hamiltonian reads,
\begin{eqnarray}
\ch_{R} &=& \ch_{0} + \ch_{SB} = (\ch_{S} + \ch_{B}) + \ch_{SB}  , \label{eqn:1} \\
\ch_{S} &\equiv& E_{e} \ee, \label{eqn:2} \\
\ch_{B} &\equiv& \hbar \sum_{k} \epsilon_{k} b_{k}^{\dagger}b_{k}, \label{eqn:3} \\  
\ch_{SB} &\equiv& \hbar \ee \sum_{k} h_{k} \epsilon_{k}  (b_{k}+ b_{k}^{\dagger}). \label{eqn:4}
\end{eqnarray} 

In order to suppress the decoherence, we apply pulses that are sufficiently short and strong.  This indicates that the interaction with the reservoir is neglected during pulse application:
\begin{eqnarray}
\ch_{SP}(t) &=& \ch_{S} + \sum_{j=0}^{N} \ch_{P,j}(t), \label{eqn:5} \\
\ch_{P,j}(t) &=& -\frac{1}{2} \ev_{j}(t) \cdot \muv \; ( \eg e^{-i \omega t}+ \ge e^{i \omega t}), \label{eqn:6} 
\end{eqnarray} 
where \(\ev_{j}(t)\) is the \(j\)-th applied pulse of external field.  We assume the pulse to be on resonance with the two-level system, which means \(E_{e}=\hbar \omega\).

When we apply \(N\) pulses with a pulse interval \(\tau_{s}\) and pulse duration \(\Delta t\), the time evolution of the density operator \(\rho(t)\) of the total system, is given by
\begin{eqnarray}
 \rho(t) &=& e^{-i L_{R}(t-(N\tau_{s}+\Delta t))} T_{+} [e^{-i \int_{ N \tau_{s} }^{N \tau_{s} +\Delta t} dt' L_{P,j} (t')}] \nonumber \\
&&\hspace{-2cm}  \times \{\prod_{j=0}^{N-1} e^{-i L_{R} (\tau_{s}-\Delta t)} 
T_{+} [e^{-i \int_{ j \tau_{s} }^{j \tau_{s} +\Delta t} dt' L_{P,j} (t')}]\} 
\rho(0) , \nonumber \\
\label{eqn:7}
\end{eqnarray}
where \(T_{+}\) is the time ordering symbol from right to left and \( L_{P,j}\) (\( L_{R}\) ) indicates the Liouville operator during the \(j\)-th pulse (the interaction with the reservoir) which is defined as
\begin{equation}
i L_{\nu} \;\; \cdots \;\; \equiv  \ih [\ch_{\nu} \;\; , \cdots\;\;],  \;\; (\nu=\{P,j\} \; or \; \{R\}). 
\label{eqn:8}
\end{equation}  

We rewrite Eq.(\ref{eqn:7}) by using the following relation for an arbitrary operator \(X\),
\begin{equation}
 e^{-i L_{\nu} (t-t_0)} X = e^{-\frac{i}{\hbar} H_{\nu} (t-t_0)} X e^{\frac{i}{\hbar} H_{\nu} (t-t_0)},
\label{eqn:9}
\end{equation}  
which consists of the operators as
\begin{eqnarray}
e^{-\frac{i}{\hbar}  H_{R} (t-t_0)} &=& e^{- \frac{i}{\hbar} H_{0} (t-t_0)} \; T_{+} [e^{-\frac{i}{\hbar} \int_{0}^{t-t_0} dt' {\tilde H_{SB}}(t')}] \nonumber \\
&=& ( U_{1}(t-t_0) \ee + U_{2}(t-t_0) \gg) \label{eqn:10}\\
T_{+} [e^{-\frac{i}{\hbar} \int_{t_0}^{t} dt' H_{P,j}(t)}] &=& e^{- \frac{i}{\hbar} H_{0} (t-t_0)} e^{- \frac{i}{\hbar} {\hat H_{P,j}} (t-t_0)} \nonumber \\
&=& U_{2}(t-t_0) \cos(\frac{\theta_{j}}{2})(e^{-i \omega (t-t_0)} \ee+\gg) \nonumber \\
&& -i \sin(\frac{\theta_{j}}{2})(\eg+\ge),  
\label{eqn:11}
\end{eqnarray} 
where
\begin{eqnarray}
 {\tilde H_{SB}}(t)&=&e^{-\frac{i}{\hbar} H_{0} t} H_{SB} e^{\frac{i}{\hbar} H_{0} t}
  =\sum_{k} h_{k} \epsilon_{k}  (b_{k} e^{-i\epsilon_{k} t} + b_{k}^{\dagger}e^{i\epsilon_{k} t}),\\
{\tilde H_{P,j}}(t)&=&e^{-\frac{i}{\hbar} H_{0} t} H_{P,j} e^{\frac{i}{\hbar} H_{0} t}
=-\frac{1}{2} \ev_{j}\cdot \muv \; ( \eg + \ge) \equiv {\hat H_{P,j}}. 
\end{eqnarray} 
Here we have assumed the each applied pulse to be square whose strength is \(E_{j}\), which gives pulse area \(\theta_{j}= \frac{\ev_{j} \cdot \muv}{\hbar} (t-t_0)\) for the \(j\)-th pulse.  In  Eq.(\ref{eqn:11}), we used the following definitions as, 
\begin{eqnarray}
U_{1}(t) &=& \exp[-i(\omega + \sum_{k} \epsilon_{k} b_{k}^{\dagger}b_{k}) t ] \;T_{+}[\exp[-i \int_{0}^{t} dt' \sum_{k} h_{k} \epsilon_{k}  (b_{k} e^{-i\epsilon_{k} t'} + b_{k}^{\dagger}e^{i\epsilon_{k} t'} )]] \nonumber \\
&=&\eta(t)\exp[-i(\omega + \sum_{k} \epsilon_{k} b_{k}^{\dagger}b_{k})t ] \exp[\sum_{k} \epsilon_{k} 
( b_{k}^{\dagger} \xi_{k} (t) - b_{k} \xi_{k}^{*} (t) )], \label{eqn:12} \\
U_{2}(t) &=& \exp[-i \sum_{k} \epsilon_{k} b_{k}^{\dagger}b_{k} t  ], \label{eqn:13}
\end{eqnarray} 
where
\begin{equation}
\eta(t)=\exp[ i \sum_{k} h_{k}^2 (\epsilon_{k} t-\sin \epsilon_{k} t)], \;\; 
\xi_{k} (t) = \frac{h_{k}}{\epsilon_{k}} (1-e^{i \epsilon_{k}  t}).   
\label{eqn:14} 
\end{equation}

Now we suppose the pulse area \(\theta_{j}\) to be \(\pi\) except for the first pulse whose pulse area is \(\frac{\pi}{2}\) to generate a superposed two-level state at an initial time (\(t=0\)).  
Defining the intensity of off diagonal element of the density operator \(\rho(t)\) as
\begin{equation}
I(t) = |Tr_{R} \langle e | \rho(t) | g \rangle|^2 , \label{eqn:17}
\end{equation}
where \(Tr_{R} \) denotes the operation to trace over the reservoir variable,
we obtain for even \(N\),
\begin{equation}
I(t)=  |Tr_{R}  [U_{1}(t-N \tau_{s}) (U_{2}(\tau_{s}) U_{1}(\tau_{s}))^{N/2} \rho(0) (U_{2}^{\dagger}(\tau_{s}) U_{1}^{\dagger}(\tau_{s}))^{N/2} U_{2}^{\dagger}(t-N \tau_{s})] |^2 , \label{eqn:15}
\end{equation}  
and for odd \(N\),
\begin{equation}
I(t)=  |Tr_{R}  [U_{1}(t-N \tau_{s}) (U_{2}(\tau_{s}) U_{1}(\tau_{s}))^{(N-1)/2} U_{2}(\tau_{s}) \rho(0)  U_{1}^{\dagger}(\tau_{s}) (U_{2}^{\dagger}(\tau_{s}) U_{1}^{\dagger}(\tau_{s}))^{(N-1)/2} U_{2}^{\dagger}(t-N \tau_{s}) ] |^2. \label{eqn:16}
\end{equation}  
Here we have assumed the pulse duration \(\Delta t\) to be infinitely small.
In the case of the pulsed magnetic resonance or the transient nonlinear optics,  \(I(t)\) indicates the signal intensity.
 
Next, we focus on the time evolution of the boson reservoir, eliminating the two-level system that periodically changes its state between \(|e\rangle\) and \(|g\rangle\) by the \(\pi\) pulse train. Denoting the displacement operator as
\begin{equation}
D(\{\alpha_{k}\}) \equiv \exp[\sum_{k} (\alpha_{k} b^{\dagger}_{k} - \alpha^{*}_{k} b_{k})], \label{eqn:18}
\end{equation}  
where \(\{\cdots\}\) means a set of bosons in the reservoir, we obtain the off diagonal element of the density operator \(\rho(t)\) in the form for even \(N\),
\begin{equation}
I(t) = |Tr_{R} \langle e| [ (|A_{N}(t) \rangle |e \rangle) ( \langle B_{N}(t)| \langle g|)]|g\rangle|^2 = |\langle B_{N}(t)|A_{N}(t) \rangle|^2,\label{eqn:19} 
\end{equation} 
with
\begin{equation}
|A_{N}(t) \rangle = D( \{ \alpha_{N,k} (t)\} ) | 0 \rangle = | \{ \alpha_{N,k} (t)\} \rangle \;\; , \;\;
\langle B_{N}(t) |  =  \langle 0 |D( \{ \beta_{N,k} (t)\} )|= \langle \{ \beta_{N,k} (t)\}|. \label{eqn:20} 
\end{equation} 
Here we defined \(\alpha_{N,k} (t)\) and \(\beta_{N,k} (t)\) as
\begin{equation}
\alpha_{N,k} (t) \equiv -h_{k}+\sum_{j=0}^{N} (-1)^{j} \{h_{k}e^{-i \epsilon_{k}(t-j\tau_{s})}\} \;\; , \;\;
\beta_{N,k} (t) \equiv \sum_{j=1}^{N}  (-1)^{j-1} \{h_{k}e^{-i \epsilon_{k}(t-j\tau_{s})}\}. \label{eqn:21} 
\end{equation} 
For odd \(N\), we obtain,
\begin{equation}
I(t) = |Tr_{R} \langle e| [ (|B_{N}(t) \rangle |e \rangle) ( \langle A_{N}(t)| \langle g|)]|g\rangle|^2 
= |\langle \{\alpha_{N,k} (t)\} | \{\beta_{N,k} (t)\} \rangle |^2,\label{eqn:22} 
\end{equation} 
with
\begin{equation}
\alpha_{N,k} (t) \equiv \sum_{j=0}^{N} (-1)^{j} \{ h_{k}e^{-i \epsilon_{k}(t-j\tau_{s})} \} \;\; ,\;\;
\beta_{N,k} (t) \equiv \sum_{j=1}^{N} (-1)^{j-1} \{h_{k}e^{-i \epsilon_{k}(t-j\tau_{s})}-h_{k} \}. \label{eqn:23} 
\end{equation} 

In obtaining Eqs.(\ref{eqn:19}) \(\sim\) (\ref{eqn:23}),  we assume that the boson reservoir is in the vacuum state and the two-level system is in the ground state at the initial time: 
\begin{equation}
\rho(0) = | g \rangle \langle g | \otimes | 0 \rangle \langle 0 |. \label{eqn:25}
\end{equation}  

Eqs. (\ref{eqn:19}) and (\ref{eqn:22}) imply that the intensity \(I(t)\) is described with overlap between coherent states \(|A_{N}(t) \rangle \) and \(|B_{N}(t) \rangle \) of the reservoir.  These coherent states are biuniquely associated to \(|g \rangle\) and \(|e \rangle\): For even \(N\), the state \(|A_{N}(t) \rangle \) is associated to \(|e \rangle\), whereas \(|B_{N}(t) \rangle \) is  with  \(|g \rangle\). The \(\pi\)-pulse application alternately exchange the association between the reservoir states and the two-level system.  

An actual evaluation of Eqs.(\ref{eqn:19}) \(\sim\) (\ref{eqn:23}) requires us to rewrite the summation over \(k\) into the energy integral, 
\begin{equation}
\sum_{k} |h_{k}|^2  f(\omega_{k}) = \sum_{k} |h_{k}|^2 f(\omega_{k})  \int_{0}^{\infty}  de \delta(e-\omega_k)  = \int_{0}^{\infty}  de h(e) f(e),  
\label{eqn:26}
\end{equation} 
where we have defined coupling spectral density \(h(e)\) as,
\begin{equation}
h(e) \equiv \sum_{k} |h_{k}|^2 \delta(e-\omega_k).  
\end{equation}

%%%%%%%%%%% sec.3 %%%%%%%%%%%
\section{Numerical evaluation}
\label{sec:3} 
%%%%%%%%%%%%%%%%%%%%%%%%%%%%%
Now we evaluate the time evolution of the intensity \(I(t)\) 
\begin{equation}
I(t) = |\langle A_{N}(t) |B_{N}(t) \rangle |^2 =\exp[-\sum_{k}|\alpha_{N,k} (t)-\beta_{N,k}(t)|^2].
\label{eqn:27}
\end{equation}
In the following, we discuss the time dependence of \(I(t)\) for non-Lorentzian and Lorentzian coupling spectral density.  While the SPC can be effective for the former case, it is ineffective for the latter case. 
 
%%%%%%%%%%% sec.3a %%%%%%%%%%%
\subsection{Non-Lorentzian coupling spectral density}
\label{sec:3a} 
%%%%%%%%%%%%%%%%%%%%%%%%%%%%%
As the first example of the non-Lorentzian coupling spectral density,  we consider a Gaussian distribution with the mean frequency \(\omega_{p}\) and the variance \(\gamma_{p}\),
\begin{equation}
h_{G}(e) \equiv \frac{s }{\sqrt{\pi} \gamma_{p} } \exp(-\frac{(e-\omega_{p})^2}{\gamma_{p}^2}).
\label{eqn:28}
\end{equation}

\begin{figure}[h]
\begin{center}
\includegraphics[scale=0.6]{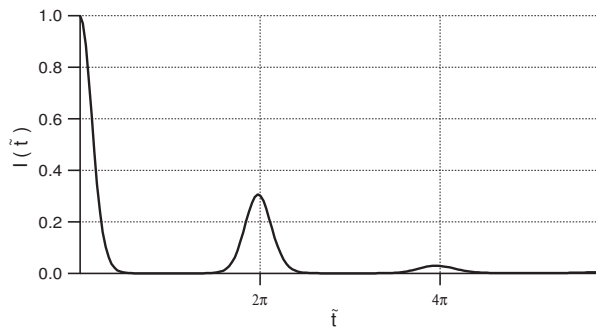}
\end{center}
\caption{Time evolution of \(I({\tilde t} )\) without pulse control for \({\tilde \gamma_p\equiv \gamma_p/\omega_p} =0.15 \), \( s= 3\).} 
\label{fig:fg1}
\end{figure}

Setting \(N=0\) in Eq.(\ref{eqn:19}), we evaluate time evolution of \(I(t)\) after a single \(\frac{\pi}{2}\) pulse at \(t=0\), which is shown in Fig.\ref{fig:fg1}.  Here and henceforth, we have used a scaled time variable as \({\tilde t \equiv \omega_{p} t} \) and set the parameters as \({\tilde \gamma_p\equiv \gamma_p/\omega_p}=0.15 \), \( s= 3\), which mean that the decay time of the interaction mode is relatively long, the average number of boson which interact with the spin is 3. We see a damped oscillation whose period is \( 2 \pi\).
%%%%%%%%%%%%%%% fig.1 %%%%%%%%%%%%%%%%%%% 
%%%%%%%%%%%%%%% fig.1 %%%%%%%%%%%%%%%%%%%

\begin{figure}[h]
\begin{center}
\includegraphics[scale=0.6]{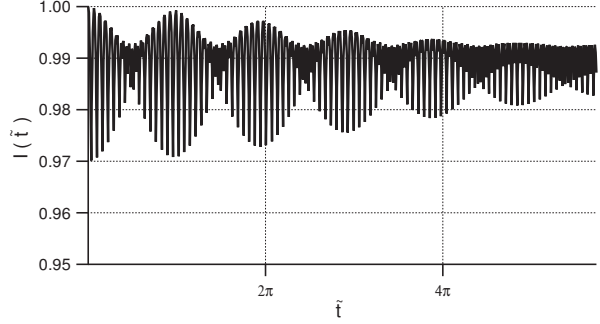}
\end{center}
\caption{Time evolution of \(I({\tilde t} )\) for the pulse interval \(\tau_s =\frac{\tau_p}{20 \pi}\). Other parameters are same as in Fig.\ref{fig:fg1}.}
\label{fig:fg2}
\end{figure}

The dynamical decoupling method\cite{lloyd1,lloyda,lloydb} tells us that an application of \(\pi\) pulse-train is sufficiently effective when the pulses are applied in a sufficiently ``small" interval. When the spectral density has a cutoff frequency at \( \omega_{c} \) as assumed in \cite{lloyda,lloydb}, the pulse interval \(\tau_{s}\) is required to be much smaller than \(\tau_{c} \equiv \omega_{c}^{-1}\) in order to control the decoherence.  In the case of the Gaussian distribution as Eq.(\ref{eqn:28}) with relatively small variance, we suppose that the pulse interval is required to be much smaller than \( \omega_{p}^{-1} \equiv \frac{\tau_{p}}{2\pi} \) for decoherence control.  In Fig.\ref{fig:fg2},  we can see that the decay of \(I({\tilde t} )\) is well suppressed for \( \tau_s =\frac{\tau_p}{20 \pi}\).
   
%%%%%%%%%%%%%%% fig.2 %%%%%%%%%%%%%%%%%%%
%%%%%%%%%%%%%%% fig.2 %%%%%%%%%%%%%%%%%%%

\begin{figure}[h]
\begin{center}
\includegraphics[scale=0.6]{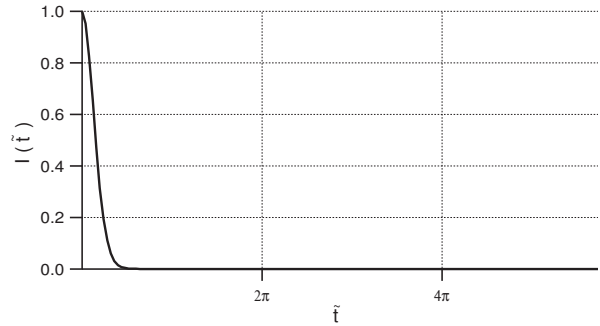}
\end{center}
\caption{Time evolution of \(I({\tilde t} )\) for the pulse interval \(\tau_s =\frac{\tau_p}{2}\). Other parameters are same as in Fig.\ref{fig:fg1}.}
\label{fig:fg3}
\end{figure}

\begin{figure}[h]
\begin{center}
\includegraphics[scale=0.6]{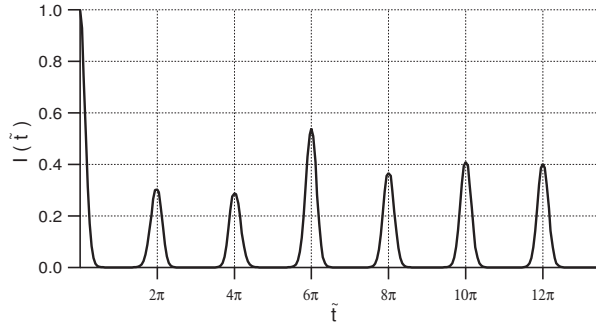}
\end{center}
\caption{Time evolution of \(I({\tilde t} )\) for the pulse interval \(\tau_s =\tau_p\). Other parameters are same as in Fig.\ref{fig:fg1}.}

\label{fig:fg4}
\end{figure}

However, when the pulse interval becomes longer, we find that the pulse application makes things even worse than the damped oscillation in Fig.\ref{fig:fg1}. This is shown in Fig.\ref{fig:fg3} where the pulse interval is \(\tau_s=\frac{\tau_{p}}{2}  \).
%%%%%%%%%%%%%%% fig.3 %%%%%%%%%%%%%%%%%%%
%%%%%%%%%%%%%%% fig.3 %%%%%%%%%%%%%%%%%%%
Now we plot the case where \(\pi\) pulses are applied with the interval \( \tau_s=\tau_p\) in Fig.\ref{fig:fg4}, where we find that the phase coherence recovers at the pulse application time.  The peak value asymptotically goes to be constant, which reflects that the dephasing in long time region cannot be described by the linear interaction. When we obtain a recovery of the intensity by synchronizing the pulse application with the characteristic period \(\tau_p\), we call this strategy for suppression of decoherence as synchronized pulse control. In this paper, we consider only the linear interaction between the spin and the original boson reservoir.  While it is necessary to take into account the nonlinear interaction, which causes the pure dephasing phenomena (irreversible processes in the long time region) in many systems\cite{uchiyama}, the effect of the pure dephasing is not significant in the time region shown in Fig.\ref{fig:fg4}, since the nonlinear interaction is often much weaker than the linear one. 
  
%%  
%%%%%%%%%%%%%%% fig.4 %%%%%%%%%%%%%%%%%%%
%%%%%%%%%%%%%%% fig.4 %%%%%%%%%%%%%%%%%%%

The synchronized pulse application has been discussed in the context of transient optical nonlinear spectroscopy, called synchronized quantum-beat echoes(SQBE) by Tanigawa, et.al.\cite{tanigawa}.  They used two light pulse trains in order to cause an optical transition between a ground-state sublevel pair. The repetition frequency of the pulse trains is equal to the separation of the sublevel pair, which is superficially similar to our approach.  However, the role of the pulse train in this paper is essentially different from SQBE, because each pulse area in SQBE is much smaller than \(\pi\). In the pulse train control of decoherence in the present work, it is essential that the pulse area of each pulse is \(\pi\), because the physical origin of the coherence recovery is the time reversal operation caused by {\it each} \(\pi\) pulse.  In Tanigawa's work, the maximum signal is generated when the total pulse area of the second pulse train is \(\pi\), and the pulse train is used to achieve the sublevel resonance. 

Next, we assume the coupling spectral density to have semi-elliptic distribution,
\begin{equation}
h_{S}(e) \equiv s \frac{1}{p}\sqrt{-(e-\omega_{p})^2+p}, 
%hr->p,vr->q
\label{eqn:29}
\end{equation}
defining \(p\) as
\begin{equation}
p \equiv \frac{4 \gamma_{p}^2}{3} \label{eqn:30}
\end{equation}
to have half width \(\gamma_{p}\). 
The coupling function has been used to describe the coupling strength between phonons and a localized electron in a solid\cite{toyozawa1}.  

\begin{figure}[h]
\begin{center}
\includegraphics[scale=0.6]{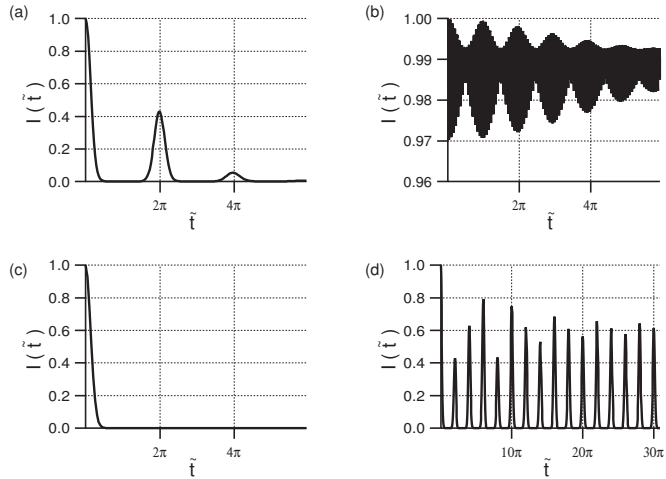}
\end{center}
\caption{Time evolution of \(I({\tilde t} )\) for semi-elliptic coupling spectral density with \(r=1\), \(s=3\), and \({\tilde \gamma_p} =0.15\); (a) without pulse application, (b) pulse interval \(\tau_s =\frac{\tau_p}{20 \pi}\), (c) pulse interval \(\tau_s =\frac{\tau_p}{2}\), (d) for pulse interval \(\tau_s =\tau_p\). }
\label{fig:fg5}
\end{figure}

We show the time evolution of  \(I(t)\) for the same parameters in Fig.\ref{fig:fg4}.  Fig.\ref{fig:fg5} shows similar behavior as in Fig.\ref{fig:fg4} except the fact that the degree of the suppression for semi-elliptic coupling spectral density is larger than the one for the Gaussian spectral density.

%%%%%%%%%%%%%%% fig.5 %%%%%%%%%%%%%%%%%%%
%%%%%%%%%%%%%%% fig.5 %%%%%%%%%%%%%%%%%%%

\begin{figure}[h]
\begin{center}
\includegraphics[scale=0.6]{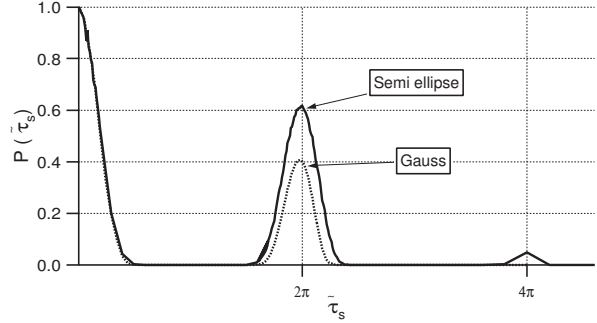}
\end{center}
\caption{Pulse-interval dependence of the asymptotic peak value \(P({\tilde \tau_s})\) for the same parameters as in Fig. \ref{fig:fg4} and \ref{fig:fg5}. The solid (dotted) line shows the \(P({\tilde \tau_s})\) for semi-elliptic (Gaussian) coupling spectral density.  }
\label{fig:fg6}
\end{figure}

Defining the asymptotic peak value of \(I({\tilde t} )\) for a pulse interval  \({\tilde \tau_s}\) as \(P({\tilde \tau_s} )\),  we show the pulse-interval dependence of  \(P({\tilde \tau_s} )\) in Fig.\ref{fig:fg6}.  The solid (dotted) line in Fig.\ref{fig:fg6} plots the \(P({\tilde \tau_s})\) for semi-elliptic (Gaussian) coupling spectral density.  We find that \(P({\tilde \tau_s} )\) has a local maximum when the pulse interval \({\tilde \tau_s}\) is close to  \( 2 \pi \).  Since \(P(2\pi)\) is nearly equal to \(P(1)\) as shown in Fig.\ref{fig:fg6}, we find that the same degree of decoherence suppression is obtained for much longer pulse interval by paying attention to the dynamical motion of the reservoir. However, the effectiveness of the SPC decreases with increasing of the width \(\gamma_p\) of the coupling spectral density. 

%%%%%%%%%%%%%%% fig.6 %%%%%%%%%%%%%%%%%%%
%%%%%%%%%%%%%%% fig.6 %%%%%%%%%%%%%%%%%%%

%%%%%%%%%%% sec.3b %%%%%%%%%%%
\subsection{Lorentzian coupling spectral density}
\label{sec:3b} 
%%%%%%%%%%%%%%%%%%%%%%%%%%%%%
Now we assume the coupling spectral density to have the Lorentzian distribution as,  
\begin{equation}
h_{L}(e) \equiv  \frac{s}{\pi} \frac{\gamma_{p}}{(e-\omega_{p})^2+\gamma_{p}^2}. 
\label{eqn:31}
\end{equation}

The Lorentzian coupling function has been often used in quantum optics. A relaxation process of an atomic system or quantum dots in a high-Q cavities has been described with a structured reservoir. The structure is determined by a distribution of coupling constants and often described with a Lorentzian function\cite{lai,kimble,lambropoulos}. 

\begin{figure}[h]
\begin{center}
\includegraphics[scale=0.6]{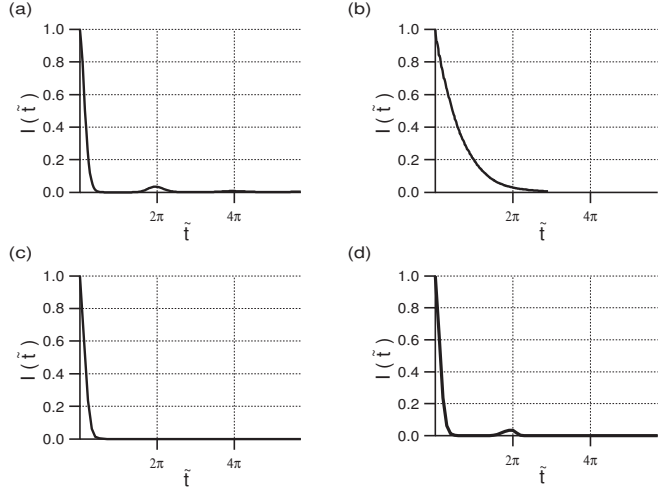}
\end{center}
\caption{Time evolution of \(I({\tilde t} )\) for Lorentzian coupling spectral density with \({\tilde \gamma_p} =0.15\); (a) without pulse application, (b) pulse interval \(\tau_s =\frac{\tau_p}{20 \pi}\), (c) pulse interval \(\tau_s =\frac{\tau_p}{2}\), (d) pulse interval \(\tau_s =\tau_p\). }
\label{fig:fg7}
\end{figure}

We show the time evolution of \(I(t)\) for \(s=3\) and \({\tilde \gamma_p} =0.15\) in Fig.\ref{fig:fg7}.  Without \(\pi\) pulse application,  \(I(t)\) shows the damped oscillation as in  Fig.\ref{fig:fg7}(a).  Contrary to the previous two cases of Gaussian and semi-elliptic distribution, Fig.\ref{fig:fg7}(b) shows that we cannot obtain the sufficient decoherence suppression for short pulse interval \(\tau_s =\frac{\tau_p}{20 \pi}\). Increasing the pulse interval to \(\tau_s =\frac{\tau_p}{2}\), we find that the degree of suppression becomes worse(Fig.\ref{fig:fg7}(c)).  The time dependence under SPC is shown in Fig.\ref{fig:fg7}(d). We find that the almost the same time evolution as the one without pulse control.  

%%%%%%%%%%%%%%% fig.7 %%%%%%%%%%%%%%%%%%%
%%%%%%%%%%%%%%% fig.7 %%%%%%%%%%%%%%%%%%%

\begin{figure}[h]
\begin{center}
\includegraphics[scale=0.6]{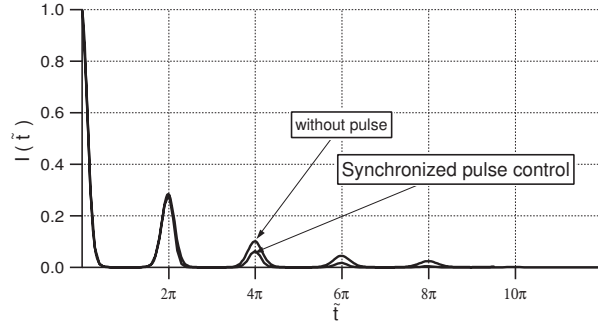}
\end{center}
\caption{Time evolution of \(I({\tilde t} )\) for Lorentzian coupling spectral density with \({\tilde \gamma_p} =0.04\). The evolution with the SPC is compared the one without pulse application. }
\label{fig:fg8}
\end{figure}

It should be noted that the ineffectiveness of the SPC does not come from the fact that the decay occurs faster for the Lorentzian coupling spectral density than the non-Lorentzian one.  In order to make the point clear, we show the time dependence for smaller \({\tilde \gamma_p}\) \(( =0.04)\), where we find the larger amplitude of oscillation in the time evolution without pulse application in Fig.\ref{fig:fg8}. The SPC makes the situation even worse, and is ineffective for the system with Lorentzian coupling spectral density. We will discuss the physical background of the ineffectiveness of the SPC in the next section. 

%%%%%%%%%%%%%%% fig.8 %%%%%%%%%%%%%%%%%%%
%%%%%%%%%%%%%%% fig.8 %%%%%%%%%%%%%%%%%%%

%%%%%%%%%%% sec.4 %%%%%%%%%%%
\section{Discussion}
\label{sec:4} 
\begin{figure}[h]
\begin{center}
\includegraphics[scale=0.6]{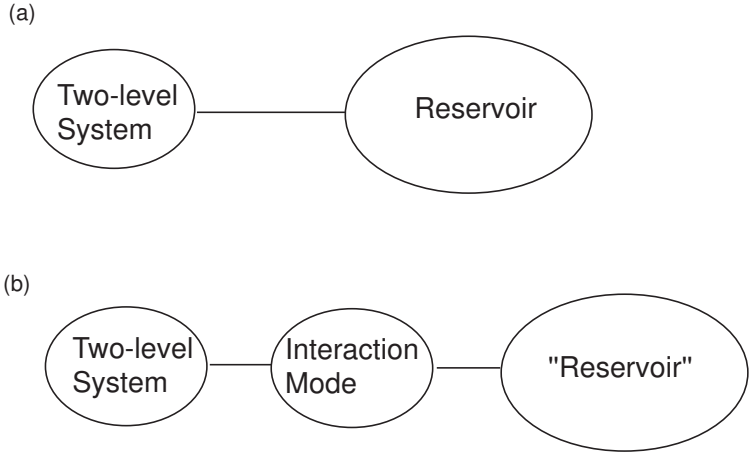}
\end{center}
\caption{Schematic representation of the two pictures for the boson system: (a) the normal mode picture, (b)the interaction mode picture. }
\label{fig:fg9}
\end{figure}

%%%%%%%%%%% sec.4 %%%%%%%%%%%
Let us consider the reason why the SPC is ineffective for the case of Lorentzian coupling spectral density.  Here we use a picture which provides us an qualitative understanding of the physical process of the SPC. The picture is obtained by replacing the original boson reservoir by a two-step structured reservoir where a single harmonic oscillator is coupled to a new ``reservoir'' with a coupling function which is different from the original one. (The ``reservoir'' means the newly introduced reservoir that causes the decay of the single harmonic oscillator.) The single harmonic oscillator is called as a quasi mode for atom-cavity system\cite{lang,barnett,dalton1,dalton2,dalton} or an interaction mode for electron-phonon system\cite{toyozawa}. In the following, we call the new harmonic oscillator as the interaction mode(Fig.\ref{fig:fg9}). The motion of the interaction mode, which is determined by the coupling to the ``reservoir'', is characterized by the original coupling spectral density.  The frequency (decay constant) of the motion of the interaction mode corresponds to the center frequency (width) of the original coupling spectral density, respectively. 

%%%%%%%%%%%%%%% fig.9 %%%%%%%%%%%%%%%%%%%
%%%%%%%%%%%%%%% fig.9 %%%%%%%%%%%%%%%%%%%

The application of a \(\pi\) pulse causes time reversal to the two-level system. Since the two-level system is coupled to the interaction mode, the degree of time reversal depends on the reversibility of the interaction mode.  In the SPC, we equalize the pulse interval to the oscillation period of the interaction mode.  If the reversibility of the interaction mode partially remains at the pulse application times, the SPC is effective for suppression of decoherence. 

When the original spin-boson interaction is characterized by the Lorentzian coupling spectral density, it has been shown that the interaction mode is coupled to the ``reservoir'' with a flat (white) coupling spectral density\cite{dalton,toyozawa1}.  The interaction mode shows the Markovian nature, which indicates the irreversibility of the motion of the interaction mode(see Appendix).  The SPC is ineffective for the Lorentzian spectral density. 

%For the case of the non-Lorentzian coupling spectral density, we can also use the two-step %structure where the interaction between the interaction mode and the ``reservoir'' is %%characterized by non-white coupling spectral density.  This implies that the time evolution of the %interaction mode is non-Markovian(partially reversible), which makes the SPC effective. In many %systems such as localized-electron phonon systems, the shape of the coupling spectral density has %clear maximum and minimum cut-off frequencies, and is usually very different from Lorentzian.  We %therefore expect that the SPC is effective enough for various systems. In order to execute the %SPC effectively, we have one more condition in addition to the non-Markovian feature of the %interaction mode.  Since the width of the coupling spectral density, \(\gamma_p\), determines the %decay constant of the interaction mode, the motion of the interaction mode decays more rapidly as %\(\gamma_p\) increases.  In many systems such as localized-electron phonon systems, the shape of %the coupling spectral density has clear maximum and minimum cut-off frequencies.  The oscillation period of the interaction mode, which is the pulse %interval in the SPC, is demanded to be smaller        

For the case of the non-Lorentzian coupling spectral density, we can also use the two-step structured reservoir where the interaction between the interaction mode and the ``reservoir'' is characterized by a non-white coupling spectral density.  This implies that the time evolution of the interaction mode is non-Markovian, and partially reversible at the pulse application time.  One should note that the non-Lorentzian coupling spectral density may not always guarantee the effectiveness of pulse control when the coupling spectral density has a slow power-law tail at high frequencies as a recent study indicates\cite{lidars}.  However, in many systems such as localized-electron phonon systems, the shape of the coupling spectral density has clear maximum and minimum cut-off frequencies.  The detailed characteristics of the spectral density for an effective pulse control deserve further research.   

We should remark that in the limit of short pulse interval, the pulse 
control is effective even for the Lorenzian coupling function.
In this case, the interaction mode does not oscillate, and the 
result is consistent with the ordinary dynamical decoupling\cite{lloyd1,lloyda,lloydb}.

%%%%%%%%%%% sec.5 %%%%%%%%%%%
\section{Concluding Remarks}
\label{sec:5} 
%%%%%%%%%%% sec.5 %%%%%%%%%%%
We have proposed a new strategy to suppress decoherence by multipulse control, which is done by synchronizing a \(\pi\) pulse train with the dynamical motion of reservoir.  We have discussed the effectiveness of SPC for the non-Lorentzian and the Lorentzian coupling spectral density.  For the former case, we find a periodic recovery of a quantum superposition at the pulse application times, whereas we cannot obtain the recovery for the latter case.    
     
Our scheme is somewhat similar to the synchronized quantum-beat echoes(SQBE)\cite{tanigawa} in the sense that pulses are applied synchronously with the dynamical feature of reservoir. However,  the SQBE is obtained by accumulating the response to each pulse whose area is much smaller than \(\pi\).  This means that the sufficient time reversal does not occur in the SQBE scheme and that the obtained echo is different from ours where essential physical origin is time reversibility caused by the each \(\pi\) pulse application. 

The SPC provides us an another kind of method to suppress the decoherence by paying attention to the dynamical motion of the reservoir.  We hope that the synchronized pulse control might extend the possibility of the pulse control of decoherence.  Especially, the drastic alteration in the feature of the quantum superposition by changing the pulse interval might indicate that the pulse application plays both roles to keep and erase a quantum memory. 
\begin{acknowledgements}
This study is supported by the Grant in Aid for Scientific Research from the Ministry of Education, Science, Sports and Culture of Japan. The authors are deeply thankful to Dr. L. Viola and the anonymous reviewer for their thoughtful comments and suggestions.

\end{acknowledgements}

%%%%%%%%%%% appendix %%%%%%%%%%%
\appendix*
\section{Introduction of the two-step model}
%%%%%%%%%%%%%%%%%%%
In this appendix, we briefly review how to introduce the two-step model with using the interaction mode(or the quasi mode).  The interaction mode is the single harmonic oscillator, and the two-level system couples only to this oscillator.  The annihilation operator of the interaction mode is defined by the linear combination of those for the original normal modes as 
\begin{equation}
B \equiv g^{ - 1} \sum\limits_{k}^{} {h_{k} \varepsilon _{k} b_{k} }, \;\;
\label{eqn:d0a}
\end{equation}
where
\begin{equation}
g \equiv \left( {\sum\limits_{k}^{} {\left| {h_{k} \varepsilon _{k} } \right|^2 } } \right)^{1/2}. \label{eqn:d0b}
\end{equation}
Then, Eq.(\ref{eqn:4}) can be rewritten as
\begin{equation}
\ch_{SB} = \hbar \ee \sum_{k} h_{k} \epsilon_{k}  (b_{k}+ b_{k}^{\dagger}) =\hbar g \ee (B+B^{\dagger}).
\label{eqn:d1}
\end{equation}
The new ``reservoir'' modes are determined to be orthogonal to the interaction mode, and the many ``reservoir'' modes are orthogonal with each other.  These boson systems other than the interaction mode do not couple to the spin system, and are called the ``reservoir'' modes.  We should note that the ``reservoir'' here implies the environmental degrees of freedom for the interaction mode, and not for the spin system.  The interaction mode and the ``reservoir'' are coupled to each other as
\begin{equation}
\ch_{BR} =\hbar (B \sum g_{j} R_{j}^{\dagger} +B^{\dagger} \sum g_{j}^{*} R_{j}), 
\label{eqn:d2}
\end{equation}
where \(R_{j}\) (\(R_{j}^{\dagger}\)) is  the annihilation (creation) operator of the oscillator for ``reservoir'' modes.  The transformation from the original normal modes to the interaction and ``reservoir'' modes provides a new picture where a two-level system interacts with a single harmonic oscillator which contacts with a ``reservoir''. 

When we consider a subsystem which consists of  the interaction mode and the ``reservoir'' with \(T=0\), the Markovian master equation which describes time evolution of the interaction mode is solved to give the exponential decay of the coherent state amplitudes\cite{phoenix}, whose decay constant is determined by the coupling spectral density \(g_{j}\).  The center frequency of the coupling spectral density in the original normal modes corresponds to the frequency of the interaction mode, and the width of the coupling spectral density is associated with the decay of the interaction mode arising from the coupling between the interaction mode and the reservoir. The Lorentzian coupling spectral density expressed by Eq. (\ref{eqn:31}) implies a Markovian time evolution of the interaction mode, which indicates the irreversibility of the motion of the interaction mode. 

The quasi mode in the quantum optics has been obtained by an analogous procedure\cite{lang,barnett,dalton1,dalton2}. We have another example to show that the two-step model with the white ``reservoir''  is equivalent to the normal mode picture with the Lorentzian coupling function\cite{hamano}.  Other than these examples, the two-step model has also been used to describe an effect of coupling between the nuclear reaction coordinate and the other coordinates on electron transfer in biomolecules\cite{garg, vitali2}.

%
% References
%

\end{document}